\documentclass[prd,aps,12pt]{revtex4}
\usepackage{psfig}
\usepackage{verbatim}
\usepackage{color}
\usepackage{rotate}
\usepackage[dvips]{graphics}
\usepackage{epsf}
\usepackage{epsfig}
\usepackage{graphicx}
\usepackage{latexsym}
\usepackage{amscd}
\usepackage{amssymb}

\newcommand{\beq}{\begin{equation}}
\newcommand{\eeq}{\end{equation}}
\newcommand{\beqn}{\begin{eqnarray}}
\newcommand{\eeqn}{\end{eqnarray}}
\newcommand\noi{\noindent}
\newcommand\la{\langle}
\newcommand\ra{\rangle}
\newcommand\eps\varepsilon

\newcommand\imag{{\imath}}

\def\fm{\,\mbox{fm}}
\def\GeV{\,\mbox{GeV}}

\def\lsim{\mathrel{\rlap{\lower4pt\hbox{\hskip1pt$\sim$}}
    \raise1pt\hbox{$<$}}}
\def\gsim{\mathrel{\rlap{\lower4pt\hbox{\hskip1pt$\sim$}}
    \raise1pt\hbox{$>$}}}

\makeatletter
\def\fmslash{\@ifnextchar[{\fmsl@sh}{\fmsl@sh[0mu]}}
\def\fmsl@sh[#1]#2{%
  \mathchoice
    {\@fmsl@sh\displaystyle{#1}{#2}}%
    {\@fmsl@sh\textstyle{#1}{#2}}%
    {\@fmsl@sh\scriptstyle{#1}{#2}}%
    {\@fmsl@sh\scriptscriptstyle{#1}{#2}}}
\def\@fmsl@sh#1#2#3{\m@th\ooalign{$\hfil#1\mkern#2/\hfil$\crcr$#1#3$}}
\makeatother

\begin{document}

\hfill{SLAC-PUB-11539}


\title{Scaling properties of high $p_T$ inclusive hadron production}

\author{S.J.~Brodsky$^1$, 
H.J.~Pirner$^2$ 
and J.~Raufeisen$^2$
}
\medskip

\affiliation{$^1$Stanford Linear Accelerator Center, Stanford University, 2575 Sand Hill Road, Menlo Park,
CA 94025, USA\\
$^2$Institut f\"ur Theoretische Physik der Universit\"at, Philosophenweg 19, 69120 Heidelberg, Germany}



\begin{abstract}
\noi
We analyze the scaling properties of inclusive hadron production in
proton-proton and in heavy ion collisions from fixed target to
collider energies. At large transverse momentum $p_T$, the invariant
cross section exhibits a power-like behavior $Ed^3\sigma/d^3p\propto
p_T^{-n}$ at  fixed transverse $x$, $x_T=\frac{2|\vec
p_{T}|}{\sqrt{s}}$, and fixed center-of-mass scattering angle
$\theta_{cm}$. Knowledge of the exponent $n$ allows one to draw
conclusions about the production mechanisms of hadrons, which are
poorly known, even at high $p_T$. We find that high-$p_T$ hadrons
are produced by different mechanisms at fixed-target and collider
energies. For pions, higher-twist subprocesses where the pion is
produced directly  dominate at fixed target energy, while
leading-twist partonic scattering plus fragmentation is the most
important mechanism at collider energies. High-$p_T$ baryons on the
other hand appear to be produced by higher-twist mechanisms at all
available energies.  The higher-twist mechanism
of direct proton production can be verified experimentally by
testing whether high $p_T$ protons are produced as single hadrons
without accompanying secondaries. In addition, we find that  
medium-induced gluon radiation in
heavy ion collisions can violate scaling. 
\\
\noi
\end{abstract}
\maketitle

\section{Introduction}

A fundamental feature of quantum chromodynamics (QCD) and asymptotic freedom is the nearly scale-invariant behavior of
quark and gluon two-particle hard-scattering processes.
If these pointlike hard-scattering subprocesses are convoluted with
the structure functions of the incident hadrons and the fragmentation functions
which produce final state interactions, the resulting inclusive cross section
$E\frac{d^3\sigma(h_ah_b\to hX)}{d^3p}$ scales nominally as $p_T^{-4}$ at  fixed
$x_R=\frac{2|\vec p_{cm}|}{\sqrt{s}}$ and $\theta_{cm}$ 
{(We denote the center-of-mass ({\em cm.})
scattering angle by $\theta_{cm}$ and the three momentum of the produced hadron in the {\em cm.} frame by $\vec p_{cm}$.
The {\em cm.} energy is $\sqrt{s}$)}.
In order to validate this scaling at $\theta_{cm}=90^\circ$, one needs measurements at various {\em cm.} energies
and corresponding values of $\vec p_{cm}$.
The conformal scaling prediction is modified by the logarithmic running of the QCD
coupling and the logarithmic corrections to the scale-free parton model
arising from the evolution of the structure functions and fragmentation functions.
These results are rigorous consequence of perturbative QCD factorization theorems
for inclusive hadron reactions at large $p^2_T$.

In the same way as Bjorken scaling provides evidence for quarks in
deep inelastic lepton scattering, the scaling behavior of the cross
sections for the production of high $p_T$ particles in hadronic
collisions can be used to test the scaling of the underlying QCD
subprocesses as well as the existence of point-like constituents
each carrying a finite fraction of the hadron's momentum
\cite{BBG,FFF}. However, in some cases such as hard exclusive
reactions, the underlying hard subprocess  cannot be the simple
$2\to 2$ reactions; since all of the valence quarks of the
interacting hadrons are involved in the scattering process.   Even
in inclusive reactions, higher parton number processes can
contribute, particularly at high $x_R$ where there is a trade-off
between the fall-off at high $p_T$ which favors minimal number of
scattering partons and the fall-off at $(1-x_R) \to 0$ which favors
more valence partons entering the hard subprocess.

It is clearly important to carefully analyze the scaling behavior of
inclusive reactions in order to confirm the validity and applicable
kinematic range of the leading-twist $2 \to 2$ subprocesses.
Our aim is to determine the mass dimension of the microscopic matrix
element from dimensional and spectator counting
rules~\cite{BF,Farrar:1975yb,Brodsky:1994kg}, which have recently
been derived nonperturbatively for nearly-conformal theories using
AdS/CFT duality~\cite{Polchinski:2001tt,BT}.

Since the dimension of this matrix element is related to the number
of participating elementary fields, such an analysis provides
detailed information about the specific microscopic hard process
underlying high $p_T$ hadron production. This is particularly
important for the interpretation of hard processes in nuclear
collisions. 

\section{Dimensional counting rules}

The partonic $S$ matrix element is related to the partonic amplitude by
$S_{fi}=\delta_{fi}+\imag(2\pi)^4\delta^{(4)}(\sum p_{in}-\sum p_{out})A_{fi}$.
With single-particle states normalized to $\la p|p^\prime\ra=2E_p(2\pi)^3\delta^{(3)}(\vec p-\vec p^\prime)$,
the amplitude $A_{fi}$ on the microscopic level
has dimension $mass^{4-n_{active}}$. Here $n_{active}$
is the number of active fields, {\em i.e.}
the number of elementary fields entering $A_{fi}$.
The simplest example is $2\to2$ scattering between quarks and gluons. In this case, $n_{active}=4$
and the partonic matrix element is dimensionless, as is natural for a scale invariant theory.
However, because of color confinement the partonic subprocess cannot be observed directly, and
one needs to find a way to connect to the hadronic reaction.

The kinematics of an  inclusive reaction $h_ah_b\to hX$
is described by 
3 Lorentz invariants 
These are {\em e.g.} the center-of-mass energy squared
$s=(P_a+P_b)^2$,
the transverse momentum transfer (squared) $t=(P_h-P_a)^2$ and the missing mass $M_X$.
It is common to introduce the dimensionless variables 
($u=M_X^2-s-t$),
\beq
x_1=-\frac{u}{s}, \quad
x_2=-\frac{t}{s}.
\eeq
These variables are related to the rapidity $y$ and radial  $x_R$ of the observed hadron by
\beqn
y&=&\frac{1}{2}\log\left(\frac{x_1}{x_2}\right),\\
\label{eq:xr}
x_R&=&\frac{2|\vec p_{cm}|}{\sqrt{s}}=1-M^2_X/s\approx x_1+x_2,
\eeqn
Since most of existing data are at $y=0$ where $x_R=x_T=2p_T/\sqrt{s}$, one often refers to
the scaling of the invariant cross section as ``$x_T$ scaling''.
For $y\neq0$, we find the variable $x_R$ more useful than $x_T$, since $x_R$ allows a smooth matching of inclusive
and exclusive reactions in the limit $x_R\to1$.

We shall assume that at high $p_T$, the inclusive cross section takes a factorized
form, even if the microscopic mechanism is higher twist,
\beq\label{eq:fact}
d\sigma(h_ah_b\to hX)=\sum_{abc}G_{a/h_a}(x_a)G_{b/h_b}(x_b)dx_adx_b
\frac{1}{2\hat s}\left|A_{fi}\right|^2dX_fD_{h/c}(z_c)dz_c.
\eeq
The dimensionless
functions $G_{a/h_a}(x_a)$ describe the momentum distributions of partons of type $a$ in
hadron $h_a$, where $a$ may stand for quarks and gluons as well as for composite degrees of freedom, such as
diquarks and intrinsic hadrons.
These functions cannot be calculated perturbatively, except in the limits $x_{a,b}\to 1$.
For quarks and gluons, the scale dependence of the distribution functions
is
described by the DGLAP evolution equations, but the evolution of
color-neutral degrees of freedom is suppressed by at least one
power of the hard scale, since gluon radiation
off color neutral objects is suppressed.
Similar observations can be made for the fragmentation function $D_{h/c}(z_c)$, which
accounts for the transition of a parton $c$ into a hadron $h$ with momentum fraction $z_c=x_1/x_a+x_2/x_b$.
The amplitude of the hard subprocess $A_{fi}$ is assumed to be calculable in perturbative QCD.
Integration and summation over all unobserved variables, such as
the phase space $dX_f$ of the final state, is understood.

By keeping all ratios of Mandelstam variables fixed, the $x$
dependence of the distribution functions does not affect the scaling
behavior of the hadronic cross section. The factorization hypothesis
Eq.~(\ref{eq:fact}) then yields the power law
\beq\label{eq:scale2}
E\frac{d^3\sigma(h_ah_b\to hX)}{d^3p}=\frac{f(t/s,u/s)}{s^{n_{active}-2}},
\eeq
which reflects the mass dimension of the microscopic amplitude.
Hence, the inclusive cross section multiplied by $p_T^n$
with
\beq
n=2n_{active}-4,
\eeq
is a function of the dimensionless variables $y$ and $x_R$ only,
\beq\label{eq:scale}
E\frac{d^3\sigma(h_ah_b\to hX)}{d^3p}=\frac{F(y,x_R)}{p_T^{n(y,x_R)}}.
\eeq
This is the desired relation: the $p_T$ dependence of the inclusive cross section 
is directly related to the number of participants $n_{active}$
in the microscopic matrix element. In higher twist processes,
the function $F(y,x_R)$ also depends on the hadron distribution
amplitudes, which have dimension mass for mesons and dimension mass squared for baryons.
For example, the pion distribution amplitude is normalized to $f_{\pi^+}=130$ MeV.
The large-$p_T$ behavior of Eq.~(\ref{eq:scale}) is determined by the minimum number of active partons associated
with the reaction $h_a+h_b\to h+X$.
Of course, $n$ will depend on $y$ and $x_R$.
Let us illustrate the scaling behavior by the following two examples:

\begin{itemize}

\item{For $pp\to pX$, one can have the subprocess $uu\to uud\bar d$, which has $n_{active}=6$
rather than 4 active elementary fields. Because of proton
compositeness, the inclusive cross section $E\frac{d^3\sigma(p p \to
p X)}{d^3p}$ scales nominally as $f^2_N\over p_T^{8}$ at  fixed
$x_R$ and $y$ where the dimensional factor $f_N$ reflects the
physics of the proton distribution amplitude. In this process, the
proton is made directly in the short distance reaction rather than
from quark fragmentation or resonance decay. 
Because no energy is wasted in the fragmentation
process, the cross section falls off relatively slowly $\sim
(1-x_R)^7$ at large $x_R.$ In general, one finds the power law
behavior $(1-x_R)^{2 n_s -1}$ of the inclusive cross section at
large $x_R$, where $n_s$ is the total minimum number of  spectators
in $h_a,h_b$ and $h$, which do not participate in the hard
scattering reaction.  In this example $n_s = 2 + 2  = 4$. [We ignore
here complications from parton spin.]}
\item{
In the limit $x_R\to x_R^{max}\approx 1$, the missing mass $M_X$
reaches its minimum allowed value and the reaction becomes
exclusive. In this case, there are no spectator fields, so that the
number of active participants $n_{active}$ attains its maximum value, {\em
e.g.} $n=2n_{active}-4=20$ for $pp\to pp$. Thus, the leading-twist pQCD
approach to high $p_T$ hadron production must fail as $x_R$
increases toward unity and the process becomes exclusive. From the
definition of $x_R$ in Eq.~(\ref{eq:xr}) it becomes clear that this
is the case at large $x_1$ and small $x_2$ (or vice versa), {\em
i.e.} for high $p_T$ hadron production at large $|y|$. }
\end{itemize}

Finally, we remark that $x_{1,2}$ are not momentum fractions. The
latter are denoted by $x_{a,b}$. The momentum fractions are defined
only within a given parton model of hadron production, and cannot be
related to hadronic invariants. Instead, the factorization ansatz
for the cross section Eq.~(\ref{eq:fact}) leads to integrals over
$x_{a,b}$. In the case of $2\to 2$ hard scattering, the lower
integration limits are \beq\label{eq:xmin}
x_a^{min}=\frac{x_2}{1-x_1},\qquad x_b^{min}=\frac{x_2x_a}{x_a-x_1}.
\eeq In the exclusive limit $x_{1,2}\to(1\pm x_F)/2$ ($x_F$ is
Feynman $x$), both momentum fraction approach unity, so that only
valence partons are important. Hence, inclusive hadron production at
very large rapidity is unaffected by gluon saturation. Such
coherence effects disappear at the largest $x_F$. (The authors of
Ref.~\cite{breakdown} come to a similar conclusion from a different
viewpoint.) This shows that the high energy limit of QCD cannot be
completely described by the color glass condensate \cite{CGC}.
However, (nearly) exclusive reactions at $\Lambda_{QCD}\ll p_T\ll
\sqrt{s}$ still allow one to study perturbative QCD processes in a
kinematic regime where Regge theory applies.

\section{Phenomenological applications}

In real QCD
the nominal power laws discussed in the previous section receive corrections
from the breaking of scale invariance in QCD, {\em i.e.} from the running coupling and the
scale breaking of structure functions and fragmentation functions.
These corrections have been discussed a long time ago in Ref.~\cite{lecture} but have
not yet been studied quantitatively.

Including scaling violations, the inclusive cross section of Eq.~(\ref{eq:scale})
changes to
\beq\label{eq:qcd}
E\frac{d^3\sigma(h_ah_b\to hX)}{d^3p}
=\left[\frac{\alpha_s(p_T^2)}{p_T^2}\right]^{n_{active}-2}\frac{(1-x_R)^{2n_s-1+3\xi(p_T)}}{x_R^{\lambda(p_T)}}
\alpha_s^{2n_s}(k_{x_R}^2)f(y).
\eeq
The threshold behavior of the cross section  follows from spectator counting rules \cite{lecture}.
We ignore here an extra contribution to this power which arises from helicity mismatch in the fragmentation process.
The strong coupling constant $\alpha_s^{2n_s}(k_{x_R}^2)$ ($n_s$ is the number of spectator fields)
arises at large momentum fraction, since all spectators must combine their momentum to produce one high-$x$ quark.
This quark is far off-shell with virtuality
$
k_{x}^2=-\frac{k_T^2+\widetilde m_q^2}{1-x},
$
so that the high-$x$ tail of the structure function is calculable in
perturbative QCD. Here, $k_T$ is the transverse momentum of the quark and $\widetilde m_q$ is related
to the quark mass, see Ref.~\cite{lecture} for details.

Eq.~(\ref{eq:qcd}) matches smoothly onto the exclusive limit
$x_R\to 1$. This is still true in the presence of scaling violations: the correction to the simple
power $2n_s-1$ due to gluon radiation is contained in the function
\beq
\xi(p_T)=\frac{C_R}{\pi}\int_{k_{x_R}^2}^{p_T^2}\frac{dk_\perp^2}{k_\perp^2}\alpha_s(k_\perp^2)
=\frac{4C_R}{\beta_0}\ln\frac{\ln(p_T^2/\Lambda_{QCD}^2)}{\ln(k_{x_R}^2/\Lambda_{QCD}^2)}.
\eeq
Here, $\beta_0=11-2N_f/3$ is the QCD $\beta$-function,
$C_R = C_F = 4/3$ for quarks and $C_R= C_A =3$ for gluons.
Note the lower integration limit $k_{x_R}^2$:
at large $x_R$, the phase space for gluon radiation vanishes and QCD scaling violations disappear.
Hence, the simple spectator counting rules become exact at the exclusive boundary.

We shall now investigate, how QCD scaling violations affect $x_R$ scaling.
For that purpose, we define an
effective power $n_{eff}(p_T)$ by taking the logarithmic derivative
\beq\label{eq:neff}
n_{eff}(p_T)=-\frac{d\ln E\frac{d^3\sigma(h_ah_b\to hX)}{d^3p}}{d\ln(p_T)}
\eeq
of the cross section.

\begin{figure}[t]
 \centerline{
 \scalebox{1}{\includegraphics{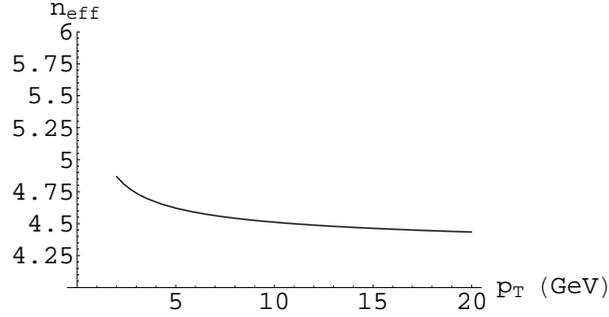}}}
    \center{
    {\caption{\em
      \label{fig:neff}The effective power $n_{eff}$ according to Eq.~(\ref{eq:neff}). The lower curve
      assumes $4$ active fields and asymptotically approaches $n_{eff}(n_{active}=4,p_T\to\infty)=4$.
      Calculations were performed at
      $x_T=0.03$ and $y=0$, which are typical values for RHIC.}
    }  }
\end{figure}

We first concentrate on RHIC kinematics at $y=0$, where rather low values of $x_T\sim 0.03$ can be reached.
Therefore, we drop all factors describing the large $x_R$ behavior of the cross section and determine $n_{eff}$
from the running coupling only. Different choices of the hard scale change numerical results by only few percent.
We also checked that the $(1-x_R)$ term is numerically irrelevant.
Results are shown in Fig.~\ref{fig:neff}.
For the lowest order process $2\to2$ process
we find
that the effective  power $n_{eff}$
approximately increases by unity.
This is close to what is seen in direct photon production at RHIC ($n_{eff}\approx5$) \cite{Barish}.

Following the suggestion of one of us (SJB), the PHENIX
collaboration has analyzed the scaling properties of data
\cite{PHENIXscaling}. For neutral pions, a value of $n_{eff}=6.33\pm
0.54$ has been determined, which is somewhat larger than the
leading-twist values shown in Fig.~\ref{fig:neff}. Strictly
speaking, the analysis of Ref.~\cite{PHENIXscaling} was done for the
invariant hadron yields rather than the cross sections. Taking into
account the variation of the inelastic cross section between
$\sqrt{s}=130$ GeV and 200 GeV, the value of $n_{eff}$ turns out to
be about 6.16, which is still within error bars. However, since
next-to-leading (NLO) order perturbative QCD is able to reproduce
RHIC data on pion production, using fragmentation functions from
$e^+e^-$ annihilation as input \cite{LT}, we conclude that pion
production is dominated by leading-twist processes.

\begin{figure}[t]
 \centerline{\scalebox{0.45}{\includegraphics{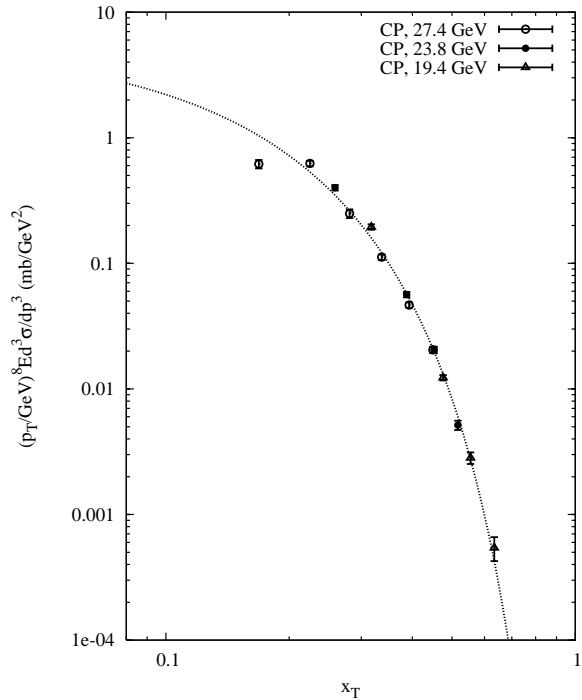}}}
    \center{
    {\caption{\em
      \label{fig:cp} Invariant cross sections for $pp\to (\pi^++\pi^-)/2+X$
      at three different energies ($\sqrt{s}=19.4$ GeV, 23.8 GeV and 27.4 GeV) multiplied by $p_T^8$.
      The power $n_{eff}= 8$ indicates
      a higher-twist mechanism.
      The curve shows the $(1-x_R)^9$ threshold behavior.}
    }  }
\end{figure}

On the other hand, the Chicago-Princeton data exhibit a strikingly different power law
and could never be described in the conventional parton model
\cite{Werner}.
An analysis
of early data on inclusive $\pi^+$ production
yields $n=8.2\pm0.5$
for $x_R=x_T\ge 0.35$,
i.e.\ $p_T\gsim 3.5$ GeV
at $\sqrt{s}\approx20$ GeV \cite{CP}.
Similar results are obtained for $\pi^-$, see Fig.~\ref{fig:cp}.
The power law $Ed^3\sigma/d^3p(pp\to\pi^+X)\propto p_T^{-8.2}$
giving
$n_{active}=6$ may indicate a quark-quark scattering
process which produces  in addition to the incoming
quarks a $q\bar q$ pair, which becomes the observed pion with high
transverse momentum.
This process has been analyzed within the Constituent Interchange Model (CIM) \cite{BBG},
where an incoming $q \bar q$ pair collides with a quark by
interchanging a quark and antiquark.
The CIM is motivated by the inclusive to exclusive transition mentioned above and is
in good agreement with the Chicago-Princeton (CP) data \cite{CP}.
The model even can reproduce the absolute normalization of the inclusive cross section.

Obviously, the production mechanism for high
$p_T$ hadrons changes from 
$\sqrt{s}=20$ GeV to 
$\sqrt{s}=200$ GeV.
For constituent interchange longitudinal momenta of O(1 GeV) can still be accommodated in the wave
function of the proton. When  the relevant longitudinal momenta are
about O(10 GeV) at higher
energies, interchange is no longer possible which
the different reaction mechanisms with increasing energy.

\begin{figure}[t]
 \centerline{\scalebox{0.45}{\includegraphics{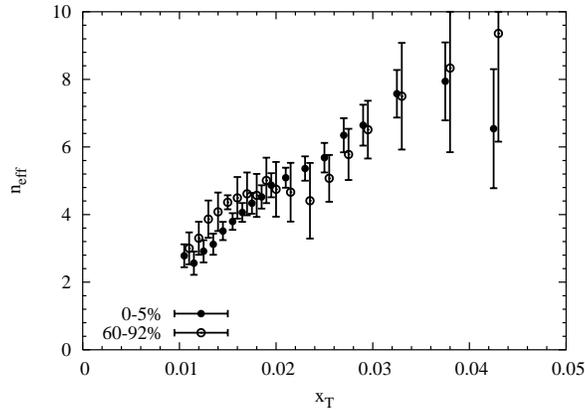}}}
    \center{
    {\caption{\em
      \label{fig:proton} Protons produced in $AuAu$ collisions at RHIC do not exhibit clear scaling properties
      in the available $p_T$ range. Shown are data for central ($0-5\%$) and for
      peripheral ($60-90\%$) collisions.}
    }  }
\end{figure}

Moreover, for proton production the $p_T$ dependence
at Chicago-Princeton energies is also explained by CIM.
A value of $n=12$ is a strong indication that higher twists from wave
function effects dominate high $p_T$
hadron production around $\sqrt{s}=20$ GeV.
Here the produced  proton is the result of proton scattering on a
quark.
If protons and pions were both produced by fragmentation
as in the Feynman-Field-Fox parton model, it is hard to understand
how a dimensionless fragmentation function could change $n$ from $8$ for pions to $12$ for protons.

Since high-$p_T$ protons are produced by higher-twist mechanisms at
fixed target energies, we also investigate the scaling properties of
proton production at RHIC. The points in Fig.~\ref{fig:proton} were
obtained from the 130 GeV data of Ref.~\cite{p130} and the 200 GeV
data of Ref.~\cite{p200}. Unfortunately, the data do not extend out
to large enough $p_T$ and error bars become too large at high $p_T$
to establish $x_T$ scaling. It is important to measure inclusive
proton production out to larger $p_T$ for at least two values of
$\sqrt{s}$. From these data one could find out whether proton
production is leading or higher twist. If protons are produced in
nuclear collisions by parton recombination (see {\em e.g.}
\cite{Fries}), the cross section should fall off exponentially, {\em
i.e.} there would be no $x_T$ scaling.

\section{Nuclear effects}

It is interesting to investigate nuclear effects on the observed
scaling laws, {\em i.e} to compare the scaling properties of
$Ed^3\sigma/d^3p(pp\to hX)$ to $Ed^3\sigma/d^3p(AA\to hX)$. Since
pions appear to be produced by a leading-twist mechanism, the
quenching of pion spectra may be due to medium induced gluon
radiation. In the following, we shall adopt the formalism of Baier
et al.\ (BDMPS-Z formalism, see Ref.~\cite{BDMPSZ} for a review).

The presence of a new dimensionful scale in nuclear collisions, namely
the BDMPS transport coefficient $\hat q$,
gives rise to the possibility that $x_T$ scaling is modified or violated.
With the quenching weight $Q(p_T)$ defined according to
\beq
\frac{1}{N_{coll}}\frac{d^2\sigma(AB\to hX)}{dydp_T^2}=\frac{d^2\sigma(pp\to hX)}{dydp_T^2}Q(p_T),
\eeq
medium effects may modify $n_{eff}$ as
\beq
n^{med}=n^{vac}-\frac{d\ln Q(p_T)}{d\ln p_T}
\eeq
The calculation of $Q(p_T)$ has been performed within the BDMPS-Z formalism in Ref.~\cite{BDMS}.
These are the basic steps:
Let $P(\Delta E)$ be the probability that a fast parton loses energy $\Delta E$ due to
gluon radiation. The medium modified cross section can then be written as the convolution
\beq\label{eq:p}
\frac{d\sigma^{med}}{d p_T^2}=
\int d(\Delta E) P(\Delta E)\frac{d\sigma^{vac}}{d p_T^2}(p_T+\Delta E).
\eeq
Since the partonic cross section is a steeply falling function of $p_T$, a small value of $\Delta E$
produces a large suppression.
For $\Delta E\ll p_T$, one can expand the logarithm of $d\sigma^{vac}/{d p_T^2}( p_T+\Delta E)$
in Eq.~(\ref{eq:p}) and obtains \cite{BDMS}
\beq\label{eq:q}
Q(p_T)\approx\int d(\Delta E) P(\Delta E)\exp\left(-\frac{\Delta E}{p_T}n^{vac}\right)
=\exp\left(-\int_0^{\omega_{max}}d\omega(1-e^{-\frac{n^{vac}\omega}{p_T}})\frac{dI}{d\omega}\right).
\eeq The advantage of this approximation is that one obtains a
particularly simple relation between $Q(p_T)$ and the gluon
multiplicity $dI/d\omega$, provided one also assumes a Poissonian
probability distribution $P(\Delta E)$. The upper integration limit
in Eq.~(\ref{eq:q}) is given by $\omega_{max}={\mathrm{min}}(\omega_{LPM},E)$, where $E=p_T$ is the energy of the fast
parton. Omitting overall numerical constants, the Landau-Pomeranchuk
effect in QCD has the gluon spectrum \beq
\frac{dI}{d\omega}\propto\alpha_s\frac{\sqrt{\omega_{LPM}}}{\omega^{3/2}},
\eeq where $\omega_{LPM}=\frac{1}{2}\hat qL^2$ ($L$ is the length of
the traversed medium) will be estimated below.

Parametrically, we obtain the following result:
depending on whether the energy $p_T$ of the fast parton exceeds the critical value
$\omega_{LPM}$,
one distinguishes the two regimes,
\beq
\Delta n=-\frac{d\ln Q(p_T)}{d\ln p_T}
\sim\left\{\begin{array}{lr}
\alpha_sn^{vac}\sqrt{\frac{\omega_{LPM}}{p_T}}&p_T\ll\omega_{LPM}\\
\alpha_sn^{vac}\frac{\omega_{LPM}}{p_T}& p_T\gg\omega_{LPM}.
\end{array}
\right.
\eeq
Hence,
$x_T$ scaling should be
strongly violated in nuclear collisions,
in contradiction to what is seen in experiment. Data are available from the PHENIX collaboration at RHIC,
see Fig.~18 of
Ref.~\cite{PHENIXscaling}.
Remarkably, despite the strong nuclear suppression of pion spectra, $n_{eff}$ for neutral pions
has almost no centrality dependence, {\em i.e.}
$n_{eff}=6.33\pm0.54$ for peripheral
and $n_{eff}=6.41\pm0.55$ for central collisions.
We conclude that radiation of a large number of soft gluons is not the dominant mechanism
behind jet quenching. 

Once again, we stress the importance of studying the cross
section at fixed $x_T$ and rapidity, rather than at fixed $\sqrt{s}$. 
In the latter case, the $p_T$ dependence of the inclusive cross section
is affected by the $x$-dependence of the structure functions, which can 
result in a $p_T$-independent quenching ratio \cite{Eskola}.
An analysis at fixed $x_T$ and $y$ eliminates the sensitivity to 
parameterizations of structure and fragmentation functions.

However, including all charged hadrons, the power $n$ increases
with centrality
from $n=6.12\pm0.49$ in $pp\to(h^++h^-)X$ to $n=7.53\pm0.44$ in $AuAu\to(h^++h^-)X$.
The reasons for this difference may be in baryon production,
since  the inclusive baryon
cross section has a steeper $p_T$ dependence.
In heavy ion collisions, pions are strongly suppressed while proton production
at not too large $p_T$ has almost no centrality dependence.
We therefore argue that  $n_{eff}=7.5$ reflects the scaling behavior of baryon production at RHIC.
The underlying mechanism could be the process $uu\to p\bar d$ with $n_{eff}=8$ as explained above.

Finally, in order to estimate $\omega_{LPM}$ (or equivalently $\hat
q$), we shall rely only on data that are not related to jet
quenching. Using Bjorken's estimate of the initial energy density,
one obtains \beq \epsilon_{Bj}=\frac{\la m_T\ra}{\pi
R_A^2\tau_0}\left(\frac{dN}{dy}\right)_{y=0} \approx 10
\GeV/\fm^3\approx 60 \epsilon_{cold} \eeq at initial time
$\tau_0=0.5$ fm. We account for the longitudinal expansion of the
medium by employing the dynamical scaling law of Salgado and
Wiedemann \cite{SW}, which relates the expanding medium to an
equivalent static scenario, \beq \hat q^{hot}=\frac{2\hat
q(\tau_0)}{L^2}\int_{\tau_0}^{\tau_0+L}d\tau
(\tau-\tau_0)\frac{\tau_0}{\tau} \approx 10\hat q^{cold}\approx
2\frac{\GeV}{\fm^2}. \eeq For $L=5$ fm we obtain $\omega_{LPM}=25$
GeV. The value of $\hat q^{cold}$ has been estimated in
\cite{relate}. Even though our estimate of $\omega_{LPM}$ may be
uncertain by a factor 3 in each direction, we are quite sure that
RHIC high $p_T$ data lie in the regime where the BDMPS-Z
medium induced gluon
radiation would yield an energy loss $dE/dz\propto
-\alpha_s\sqrt{\hat qE}$, similar to the QED Landau-Pomeranchuk
effect. A mean energy loss proportional to the projectile energy 
may be more consistent with the data \cite{Vitev}.

\section{Summary}

We have reviewed dimensional counting rules at $x_T$ scaling laws
for high $p_T$ inclusive hadron production \cite{BBG,BF}. At leading
twist, the inclusive cross section $Ed^3\sigma/d^3p$ scales
nominally as $p_T^{-n}$ with $n=4$ at fixed ratios of invariants.
Scaling violations in QCD, in particular the running coupling
constant slightly increase the value of $n$. The experimental value
$n_{eff}=6.33\pm 0.54$ for neutral pion production in peripheral
heavy ion collisions at RHIC is somewhat larger than the expectation
from leading twist. Nevertheless, we think that leading-twist
partonic scattering is dominant, since NLO perturbation theory
describes the data reasonably well \cite{LT}. At fixed  target
energies however ($\sqrt{s}\approx 20$ GeV), an effective power of
$n_{eff}\approx 8$ for pions and $n_{eff}\approx 12$ for protons is
a strong indication for higher-twist mechanisms in the SPS energy
range. The fixed target data can be reproduced in the Constituent
Interchange Model of Ref.~\cite{BBG}.

Measurements of single particle hadron 
and photon production at the new hadron facilities at GSI and J-PARC 
will be very sensitive to higher twist effects. We have shown
how one bridges the high and low energy domains. In addition, our 
analysis can be used to properly obtain the exclusive-inclusive 
connection. The conventional approach overestimates evolution at 
the exclusive limit. 

We have also investigated how medium induced gluon radiation changes
the scaling properties of high-$p_T$ hadron production within the
BDMPS-Z formalism. We find that radiation of an infinite number of
soft gluons would violate $x_T$ scaling. This is however not what is
seen in experiment: the scaling law of neutral pions is unaffected
by the nuclear medium (within error bars) \cite{PHENIXscaling}. We
conclude that medium-induced gluon radiation is not the mechanism
responsible for pion quenching at RHIC. This is further supported by
the fact that charm production appears to be strongly suppressed in
nuclear collisions \cite{PHENIXcharm}, {\em i.e.} there is no dead cone effect as one
would expect if quenching were due to bremsstrahlung \cite{DK}.

Including all charged hadrons, the effective power law changes from
$n_{eff}=6.12\pm 0.49$ in peripheral collision (60-80\%)to
$n_{eff}=7.53\pm 0.44$ in central collisions (0-10\%) at RHIC. We
argue that the larger value of $n_{eff}$ in central collisions is
due to a large baryon contribution to the charged hadron yield. If
protons are produced by a higher-twist mechanism such as $uu\to
p\bar d$ at not too high $p_T$, then there would be an intermediate
$p_T$ range in which protons scale with $n_{eff}\approx 8$. This
higher-twist mechanism is different from parton recombination models
\cite{Fries}, which lead to an exponential fall off. Unfortunately,
existing data on inclusive proton production at RHIC do not allow
one to draw definite conclusions at this time. Data at different
energies will be required to separate the contributing mechanisms.
In addition, the higher-twist mechanism of direct proton production
which we propose can be verified experimentally by testing whether
high $p_T$ protons are produced as single hadrons without
accompanying secondaries.  It is clearly essential for the correct
interpretation of the heavy-ion collision data that the role of the
higher-twist processes be definitively determined.

\medskip
\noindent {\bf Acknowledgments}: This work was supported by the
German BMBF under contract 06 HD 158, by the U.S. Department of
Energy under contract number DE-AC03-76SF00515 and by the Alexander
von Humboldt Foundation.

\end{document}